\documentstyle[12pt,aaspp4,epsf]{article}

\slugcomment{{\bf The Astronomical Journal}, Vol. {\bf 112}, September 1996}

\lefthead{Merritt}
\righthead{Axisymmetric Stellar Systems}

\begin{document}

\title{The Dynamical Inverse Problem for \\ Axisymmetric Stellar Systems}

\author{David Merritt}
\affil{Department of Physics and Astronomy, Rutgers University,
New Brunswick, NJ 08855}

\bigskip
\centerline{Rutgers Astrophysics Preprint Series No. 190}
\bigskip

\begin{abstract}
The standard method of modelling axisymmetric stellar systems begins
from the assumption that mass follows light.
The gravitational potential is then derived from the luminosity
distribution, and the unique two-integral distribution function $f(E,L_z^2)$
that generates the stellar density in this potential is found.
It is shown that the gravitational potential can instead be generated 
directly from the velocity data in a two-integral galaxy, 
thus allowing one to drop the assumption that mass follows light.
The two-dimensional rotational velocity field can also be recovered 
in a model-independent way.
Regularized algorithms for carrying out the inversions are 
presented and tested by application to pseudo-data from a 
family of oblate models.

\end{abstract}

\section{INTRODUCTION}

Modelling of elliptical galaxies has a long history, extending back to a 
time when it was taken for granted that elliptical galaxies were axisymmetric 
and that all of their mass could be accounted for in stars.
We now know that dark matter is a commmon component of 
galaxies, both at large (Ashman 1992) and small (Kormendy \& Richstone 1995) 
radii, and that triaxial configurations are possible ones for stellar systems 
(Schwarzschild 1979).

One goal of modern dynamical studies is accordingly to verify the presence 
of dark matter; another is to detect departures from axisymmetry.
Because triaxial models are difficult to construct, one often 
begins by looking for simpler models that are consistent
with the data.
This ``model-building'' approach is popular since it requires a 
minimum of thought about the information content of the data: one 
simply builds a model and checks whether it reproduces the 
observations.
But one can take a more sophisticated approach, and ask whether 
the data imply anything definite about the observed stellar 
system.
For instance, one might attempt to falsify the axisymmetric 
hypothesis in a model-independent way.
This ``inverse problem'' approach is more difficult but also more 
rewarding, since it leads to more secure conclusions about the 
dynamical state of the galaxy.

Beginning with Binney, Davies \& Illingworth's (1990) pioneering 
study, a standard scheme has been developed for modelling 
elliptical galaxies.
One assumes that the observed galaxy is axisymmetric and that the 
distribution of mass is known -- for instance, mass might be 
assumed to follow light.
The gravitational potential is then derived from this mass
distribution using Poisson's equation; the unique two-integral 
distribution function 
$f(E,L_z)$ that reproduces the stellar density in this 
potential (or more exactly, the even part of $f$) can also be found 
(Lynden-Bell 1962; Hunter 1975).
The odd part of $f$, which determines the rotational velocity 
field, is usually represented via some ad hoc parametrization.
One then projects this derived $f$ back into observable space 
to find the predicted kinematical variables and compares them 
with the observations.
If there is agreement, one can claim to have found an 
axisymmetric model that is consistent with the data. 

This technique has been used to reproduce the kinematical data for a few, 
well-observed elliptical galaxies (van der Marel et al. 1994; 
Dehnen 1995; Qian et al. 1995).
But it is difficult to state precisely what has been learned from 
studies like these.
Are the observed galaxies actually characterized by a constant 
$M/L$, axisymmetry, and a two-integral distribution function, as assumed?
Or might there exist models with spatially-varying $M/L$'s or
three-integral $f$'s that reproduce the data equally well?
And if no model consistent with the data can be found, 
which of the various assumptions made in the model-building has been 
violated?

A useful comparison can be made here with spherical systems (\S 2.1).
If one assumes that the gravitational potential $\Phi(r)$ of a 
spherical galaxy is known, then the observed velocity dispersion 
profile can be used to infer the unique dependence of anisotropy 
on radius (Binney \& Mamon 1982).
Alternatively, one can assume that the distribution of stellar velocities is 
everywhere isotropic, in which case the velocity dispersion profile implies 
a unique form for $\Phi(r)$ (Merritt \& Gebhardt 1994).
Assuming {\it both} isotropy {\it and} a constant $M/L$ would clearly be an 
over-determination of the problem, since the two assumptions 
together imply a unique velocity dispersion profile,
and this predicted profile would almost certainly be different 
from the observed profile -- indeed, the kinematical data would not have been 
used at all in the construction of the model.
But this is precisely what is commonly done in modelling 
axisymmetric galaxies: one assumes both isotropy in the meridional plane 
(i.e. $f=f(E,L_z)$) as well as an ad hoc form for the potential, and these 
two assumptions together imply a unique dependence of the mean 
square velocity on position over the image of the galaxy.
It is unlikely that a model constructed in this way would reproduce the
two-dimensional velocity dispersion data; and if it did not,
there would be no clear indication of which assumption had been violated.

It is clearly desirable to think more carefully about the 
information content of kinematical data in axisymmetric stellar systems.
Unlike the spherical case, the functions to be derived depend on two 
spatial variables, but the kinematical data are likewise 
two-dimensional, depending on both projected radius and position angle.
Many such data sets now exist, for stars in globular clusters 
(Meylan et al. 1995) and in galaxies (Bacon et al. 1994); emission-line 
objects around the Galactic bulge (te Linkel Hekkert et al. 1991) and 
around other galaxies (Hui et al. 1995); and galaxies in clusters 
(Colless \& Dunn 1996).
These data are most often in the form of discrete velocities, and 
in the best-studied systems, the measured velocities number in the 
hundreds or even thousands.

Here it is shown (\S 2.2) that the availability of two-dimensional 
kinematical data allows one to approach the dynamical study of 
axisymmetric stellar systems as an inverse problem rather than 
as a model-building problem.
From the single {\it Ansatz} $f=f(E,L_z)$, one can infer
the unique gravitational potential $\Phi(\varpi,z)$ that is 
consistent with the observed first and second velocity moments 
of $f$.
The rotational velocity field can likewise be obtained in a 
model-independent way.
By using the kinematical data in the construction of the model, rather 
than assuming that mass follows light, one thus arrives at the unique 
pair of functions $\{f,\Phi\}$ that are consistent with 
the two-integral hypothesis.
If this unique model can be falsified -- for instance, by 
using line-of-sight velocity distributions, proper motions, or 
some other data -- then the two-integral, axisymmetric hypothesis
will have been convincingly ruled out.
In the model-building approach, on the other hand, one can not discard the 
two-integral assumption until one has explored an effectively 
infinite number of possible forms for the potential (\S 2.3).

The numerical techniques that are required for converting the 
velocity data into a map of the potential are rather 
more subtle than the ones that have been applied up to now in 
the study of axisymmetric galaxies.
These techniques are accordingly described in some detail (\S3.1) before 
applying them to pseudo-data generated from a family of oblate 
models (\S3.2, 3.3).

Although this approach yields the unique potential consistent with 
the two-integral assumption for $f$, the true potential might be 
different if $f$ depends on a third integral.
However it would be premature to investigate three-integral 
models until one had first used the algorithms described here, or 
equivalent ones, to rule out a two-integral model.
Having done so, one could then search for the pair of functions 
$\{\Phi(\varpi,z)$, $f(E,L_z,I_3)\}$ that are 
most consistent with the data.
This is a hard problem, and even in the spherical geometry there is no
published algorithm that can extract $\Phi(r)$ and $f(E,L^2)$ in a 
completely model-independent way from the data; the most sophisticated 
such algorithms are still based on parametrized forms for the potential 
(Saha \& Merritt 1993; Merritt 1993b).

The focus here is on situations where the velocities are measured 
discretely, as would be the case for stars in a globular cluster.
Most large, kinematical data sets are of this form.

It is also assumed throughout that the observer lies in the 
equatorial plane of the observed system.
The reason for this unpleasant assumption is that
deprojection of the luminosity density becomes underdetermined if
the inclination angle is less than $\pi/2$, even if this angle is 
assumed known (Rybicki 1986; Gerhard \& Binney 1996).
One can therefore probably not hope to uniquely infer the 
dynamical state of an axisymmetric galaxy that is not viewed 
edge-on.
This fact is routinely ignored in the model-building studies but 
must be faced squarely if one wishes to solve the inverse 
problem.

Also presented here (\S 3.4) is the first regularized algorithm capable of 
deriving $f(E,L_z)$ for an axisymmetric system from its density 
and potential (the Lynden-Bell -- Hunter problem) given imperfect or 
incomplete information about those functions.

\section{THE INVERSE PROBLEM}
\subsection{Spherical Systems}

Our starting point for the axisymmetric inverse problem is the simpler spherical
problem.
A spherical stellar system has a distribution function that depends on the
orbital energy $E$ and angular momentum ${\bf L}$ (both defined here per unit
mass).
Spherical symmetry in velocity space is assumed as well; 
then $f=f(r,v)=f(E)=f[v^2/2+\Phi(r)]$, and the radial and azimuthal velocity
dispersions are everywhere equal, $\sigma_r(r)=\sigma_t(r)\equiv \sigma(r)$.

Having measured a set of discrete positions and velocities, one can compute
smooth approximations to the surface density profile $\Sigma(R)$ and the
line-of-sight velocity dispersion profile $\sigma_p(R)$.
The function $\sigma(r)$ that defines the intrinsic velocity
dispersion is then related to the observed velocity dispersions via
\begin{equation}
\Sigma(R)\sigma_p^2(R) = \int_{R^2}^{\infty} \nu(r)\sigma^2(r)
{dr^2\over\sqrt{r^2-R^2}} 
\label{sigobs}
\end{equation}
which when inverted gives
\begin{equation}
\nu(r)\sigma^2(r) = -{1\over \pi}\int_r^{\infty} {d(\Sigma\sigma_p^2)\over
dR} {dR\over\sqrt{R^2-r^2}}.
\label{sigmar}
\end{equation}
Here $\nu(r)$ is the spatial density of the kinematical sample, obtained by
deprojecting the surface density $\Sigma(R)$:
\begin{equation}
\nu(r) = -{1\over \pi}\int_r^{\infty} {d\Sigma\over dR}
{dR\over\sqrt{R^2-r^2}}.
\label{nur}
\end{equation}
The mass within $r$ and the potential follow from 
Jeans's equation in spherical symmetry:
\begin{equation}
GM(r) = r^2{d\Phi\over dr} = -r\sigma^2(r)\left({d\log\nu\over d\log r} +
{d\log\sigma^2\over d\log r}\right),
\label{massr}
\end{equation}
and the mass density is
\begin{equation}
\rho(r) = {1\over 4\pi r^2} {dM\over dr}.
\label{rhor}
\end{equation}
Finally, the unique isotropic distribution function describing the
kinematical sample is given by Eddington's formula:
\begin{equation}
f(E)={1\over \sqrt{8}\pi^2}{d\over dE}\int_E^0{d\nu\over
d\Phi}{d\Phi\over\sqrt{\Phi-E}}.
\label{fofe}
\end{equation}
Thus both $f$ and $\Phi$ follow uniquely from the observed surface density
and velocity dispersion profiles, under the sole 
assumption of spherical symmetry in velocity space.
\footnote{Strictly speaking, $\sigma_r=\sigma_t$ does not imply $f=f(E)$;
one can construct ``isotropic'' stellar systems with distribution functions 
that depend on $L^2$ as well.}
Here we are treating the stars as if they were ions in an X-ray emitting gas;
the quantity $\sigma^2(r)$ plays the role of $kT(r)/m$ in the gas, and no
assumption has been made that mass follows light.

In spite of its limitations, this approach has the following 
things to recommend it.

1. One begins from a statement of the problem that is fully determined: there
is a unique pair of functions $\{f(E),\Phi(r)\}$ that reproduce any set of
observed profiles $\{\Sigma(R),\sigma_p(R)\}$.
In practice one does not measure $\Sigma(R)$ or $\sigma_p(R)$ 
precisely,
but nonparametric algorithms can be defined which yield
estimates $\{\hat f, \hat\Phi\}$ that are suitably close to the 
true functions (Merritt \& Gebhardt 1994).

2. The dependence of $\Phi$ on $r$ is determined from the velocity data,
as it should be,
rather than being derived from the luminosity density under the
always-questionable assumption that mass follows light.

3. The validity of the single assumption -- that $f$ is spherically 
symmetry in velocity space -- can be checked.
One simply projects the derived model to find the predicted, joint
distribution of positions and line-of-sight velocities:
\begin{equation}
N(R,v_p) = \pi\int_{R^2}^{r_{max}^2(V)}
{dr^2\over\sqrt{r^2-R^2}}\int_0^{-2\Phi(r)-v_p^2}f\left[v'^2/2+v_p^2/2+\Phi(r)
\right]dv'^2. 
\end{equation}
If this function is consistent with the observed line-of-sight velocity
distribution at every radius, then the model is fully
consistent with the data (assuming that no other relevant data exist);
if not, isotropy can be ruled out, independent of
any preconceptions about the radial variation of M/L.

Often the kinematical data will be insufficient in quality or quantity to
produce a statistically significant discrepancy with the computed $N(R,v_p)$, 
because anisotropy generates only slight changes in the form of the 
line profiles and very high quality data are required to detect these 
changes (Gerhard 1993; Merritt 1993b).
Nevertheless this approach makes manifest a point that is often 
misunderstood: one can not, in principle, make inferences about 
the degree of velocity anisotropy in a spherical system
from the velocity dispersion data alone, since one can always 
construct the potential in such 
a way as to reproduce any observed $\sigma_p(R)$.

4. For some stellar systems -- e.g. globular clusters -- the assumption of
isotropy in velocity space can be physically motivated.

\subsection{Axisymmetric Systems}

Our goal is to identify a similar inverse problem for axisymmetric stellar
systems.
We wish to show that complete knowledge of the line-of-sight 
velocity moments ${\overline v}_p$ and $\overline{v_p^2}$ on the 
plane of the sky, and of the surface density $\Sigma$ of the 
kinematical population, is equivalent to knowledge of the 
two-integral distribution function $f(E,L_z)$ describing that 
population and of the potential $\Phi(\varpi,z)$ in which it 
moves -- independent of any assumptions about the relative 
distribution of mass and light.
As discussed above, it is assumed throughout that the observer lies 
in the equatorial plane of the observed system. 

The Jeans equations that relate the potential $\Phi(\varpi,z)$ of an
axisymmetric system to gradients in the velocity dispersions are
\begin{eqnarray}
\nu{\partial\Phi\over\partial z} & = &-{\partial(\nu\sigma_z^2)\over\partial z}
 - {\partial (\nu\overline{v_{\varpi}v_z}\varpi)\over\varpi\partial\varpi}, \\
\nu{\partial\Phi\over\partial\varpi} & = &
-{\partial(\nu\sigma_{\varpi}^2)\over\partial\varpi} -
{\partial(\nu\overline{v_{\varpi}v_z})\over\partial z} -
{\nu\over\varpi}\left(\sigma_{\varpi}^2-\overline{v}_{\phi}^2-\sigma_{\phi}^2
\right). 
\label{potential}
\end{eqnarray}
Here $(\varpi,z,\phi)$ are the usual cylindrical coordinates; all quantities
are assumed independent of $\phi$.

Now set $f=f(E,L_z)$. 
The velocity dispersions are then isotropic in the meridional plane, 
$\sigma=\sigma_{\varpi}=\sigma_z$, and 
the Jeans equations reduce to
\begin{eqnarray}
\nu{\partial\Phi\over\partial z} & = & -{\partial(\nu\sigma^2)\over\partial 
z} ,\\
\nu{\partial\Phi\over\partial\varpi} & = &
-{\partial(\nu\sigma^2)\over\partial\varpi} -
{\nu\over\varpi}\left(\sigma^2 - \overline{v}_{\phi}^2 - 
\sigma_{\phi}^2\right). 
\end{eqnarray}

Define Cartesian coordinates $(X,Z)$ on the plane of the sky, with $Z$ 
parallel to the symmetry axis and $Y$ along the line of sight.
We assume that $\nu(\varpi,z)$ has been derived from 
$\Sigma(X,Z)$ via the usual inversion, unique for an edge-on galaxy
(Rybicki 1986).
$\Phi(\varpi,z)$ is still unknown.
But if we multiply the first Jeans equation by 
$\partial/\partial\varpi~\nu^{-1}$ 
and the second by $\partial/\partial z~\nu^{-1}$ and equate, we find
a relation between $\sigma_{\phi}$, $\overline{v}_{\phi}$ and 
$\sigma$ that is independent of $\Phi$:
\begin{equation}
{\partial\nu\over\partial\varpi}{\partial\sigma^2\over\partial z} - 
{\partial\nu\over\partial z}{\partial\sigma^2\over\partial\varpi} +
{\nu\over\varpi}{\partial\over\partial z}\left(\sigma^2 - 
\sigma_{\phi}^2 - \overline{v}_{\phi}^2\right) = 0. 
\label{magic}
\end{equation}
This relation may be seen as specifying a unique $\sigma_{\phi}$ given 
$\sigma$ and $\overline{v}_{\phi}$, i.e.
\begin{equation}
\overline{v}_{\phi}^2(\varpi,z) + \sigma_{\phi}^2(\varpi,z) = 
\overline{v^2_{\phi}}(\varpi,z) = -\varpi\int_z^{\infty} 
\nu^{-1}(\varpi,z)\left[ 
{\partial\nu\over\partial\varpi}{\partial\sigma^2\over\partial z} - 
{\partial\nu\over\partial z}{\partial\sigma^2\over\partial\varpi}\right] 
dz + \sigma^2(\varpi,z) .
\label{magic2}
\end{equation}

Now consider the rotational velocity field.
The mean line-of-sight velocity $\overline{v}_p(X,Z)$ is related to 
the internal streaming velocity $\overline{v}_{\phi}(\varpi,z)$ 
via the projection integral
\begin{equation}
\label{sigmavy}
\Sigma(X,Z)\overline{v}_p(X,Z) =
2X\int_{X}^{\infty}\nu(\varpi,Z)\overline{v}_{\phi}(\varpi,Z)
{d\varpi\over\sqrt{\varpi^2-X^2}}
\end{equation}
(Fillmore 1986, Eq. 12).
Equation (\ref{sigmavy}) has inversion
\begin{equation}
\nu(\varpi,z)\overline{v}_{\phi}(\varpi,z) =
-\varpi{1\over\pi}\int_{\varpi}^{\infty} {d\over dX}
\left[X^{-1}\Sigma(X,z)\overline{v}_p(X,z)\right] {dX\over\sqrt
{X^2-\varpi^2}}.
\label{nuvphi}
\end{equation}
Note that $\overline{v}_{\phi}$ is completely determined by $\overline{v}_p$
independent of the two-integral assumption -- all that is required here is
the assumption of axisymmetry.
The possibility of computing the rotational velocity field via direct 
inversion of the data has been pointed out (though never implemented)
by a number of authors including Meylan \& Mayor (1986) and Merrifield (1991).

The mean square line-of-sight velocity is
\begin{equation}
\Sigma(X,Z)\overline{v_p^2} (X,Z) = 2\int_X^{\infty}\nu(\varpi,Z)
 \left[\left(1-{X^2\over\varpi^2}\right)\sigma^2(\varpi,Z) + 
{X^2\over\varpi^2}\overline{v^2_{\phi}}(\varpi,Z)\right]{\varpi
d\varpi\over\sqrt{\varpi^2-X^2}}
\label{disp}
\end{equation}
(Fillmore 1986, Eq. 16).
The integrand on the right hand side contains the {\it two} unknown 
functions $\sigma^2$ and 
$\overline{v^2_{\phi}}=\overline{v}^2_{\phi}+\sigma^2_{\phi}$, 
and at first sight does not admit of a unique inversion.
But we have a relation between $\sigma^2$ and 
$\overline{v^2_{\phi}}$, equation (\ref{magic2}).
We can therefore replace $\overline{v^2_{\phi}}$ 
in the integrand by a linear
expression that depends only on $\sigma^2$ and on $\nu$.
Although no analytic expression for the inverse relation could be 
found by the author, it will be shown below that $\sigma^2(\varpi,z)$ 
is numerically well-determined by this equation.
Having derived $\sigma^2(\varpi,z)$,
$\overline{v^2_{\phi}}(\varpi,z)$ then follows from equation (\ref{magic2}),
and $\sigma^2_{\phi} = \overline{v^2_{\phi}} - \overline{v}^2_{\phi}$.

It has thus been shown that $\overline{v}_{\phi}(\varpi,z)$, 
$\sigma_{\phi}(\varpi,z)$ and $\sigma(\varpi,z)$
are determined by the two-dimensional velocity moment data in an 
edge-on, two-integral system without any assumptions aside from 
isotropy in the meridional plane.
It follows that the potential $\Phi(\varpi,z)$ and the
mass density $\rho(\varpi,z)$ are also known: the former from 
equations (10) or (11), the latter from Poisson's equation:
\begin{equation}
\rho(\varpi,z) = {1\over 4\pi
G}\left[{1\over\varpi}{\partial\over\partial\varpi}
\left(\varpi{\partial\Phi\over\partial\varpi}\right) + {\partial^2\Phi\over
\partial z^2}\right]. 
\end{equation}
So derived, $\rho$ will of course not necessarily be 
proportional to $\nu$.

Finally, there is a unique distribution function $f(E,L_z)$ describing the
kinematical population that reproduces $\nu$ and $\overline{v}_{\phi}$ in the
derived potential.
As usual, we divide $f$ into odd and even parts with respect to $L_z$:
\begin{equation}
f(E,L_z)=f_+(E,L_z) + f_-(E,L_z), \ \ \ \ f_{\pm}(E,L_z)\equiv {1\over 2}
\left[f(E,L_z)\pm f(E,-L_z)\right].
\end{equation}
The number density is related to $f_+$ via
\begin{equation}
\nu(\varpi,z)={4\pi\over \varpi} \int_{\Phi}^0 dE
\int_0^{\varpi\sqrt{2(E-\Phi)}} f_+(E,L_z) dL_z, 
\label{fela}
\end{equation}
and the rotational velocity field is related to $f_-$ via
\begin{equation}
\nu(\varpi,z)\overline{v}_{\phi}(\varpi,z) = {4\pi\over \varpi^2}
\int_{\Phi}^0 dE \int_0^{\varpi\sqrt{2(E-\Phi)}} f_-(E,L_z) L_z dL_z.
\label{felb}
\end{equation}
These equations imply unique functions $f_+$ and $f_-$ given $\nu$,
$\overline{v}_{\phi}$ and $\Phi$ (Lynden-Bell 1962; Hunter 1975; Dejonghe 1986),
although the inverse relations are complex.
Once again, since mass is not assumed proportional to light, the population
described by this $f$ will not necessarily have the same spatial distribution
as the matter that determines the potential.

Practical algorithms for carrying out the inversions, suitable for 
real (i.e. noisy and incomplete) data, are described below.

Just as in the spherical case, the validity of the assumed relation between
$\sigma_{\varpi}$ and $\sigma_{\phi}$ can be verified by
computing the full distribution of line-of-sight velocities $N(X,Z,v_p)$ from
the model and comparing with the observed velocity distributions; 
or by carrying out the same procedure with the predicted and 
observed proper motions; etc.
If these predicted and observed distributions are significantly different, 
one can conclude that
the assumption of isotropy in the meridional plane is not valid
(or that the system is not being viewed edge-on, or is not 
axisymmetric) and proceed to investigate more general models.

Note that -- just as in the spherical case -- one can not infer 
the presence of velocity anisotropy ($\sigma_z\ne\sigma_{\varpi}$) 
based on inspection of the velocity dispersions alone, since the potential 
can always be constructed so as to reproduce the observed dispersions 
without a third integral.
Information about the dependence of $f$ on a third integral is 
contained only in the higher-order moments of the line-of-sight 
velocity distribution (or the proper motions); and conversely, 
there is no need to use that extra information unless one intends 
to investigate three-integral models -- the two-integral problem is fully 
constrained by the the first and second velocity moments of 
$f$.

\subsection{Comparison with Other Methods}

It is interesting to compare this approach to the one pioneered by 
Binney, Davies \& Illingworth (1990); their technique 
has recently formed the basis of a number of data-based studies by van der 
Marel and collaborators (van der Marel, Binney \& Davies 1990; 
van der Marel 1991; van der Marel et al. 1994).
Similar approaches to the axisymmetric inverse problem have been 
worked out by Merrifield (1991), Dejonghe (1993),  
Emsellem, Monnet \& Bacon (1994), Dehnen (1995), Kuijken (1995), 
and Qian et al. (1995).
All of these authors assume at the outset that mass follows light 
(or that the mass distribution has some other ad hoc form, often 
including a central point mass), and derive the two-dimensional 
potential from the assumed mass distribution using Poisson's equation.
The depth of the potential, or equivalently the mass-to-light ratio, 
is obtained from the virial theorem.
The observed velocities are not otherwise used in the construction of 
the model, except insofar as they constrain the form of the odd 
part of $f$, which is typically represented via some simple 
parametrization.

The advantage of this method is its simplicity: once the 
potential is specified, the even part of the distribution function 
$f_+(E,L_z)$ or any of its moments can be obtained 
(Lynden-Bell 1962; Hunter 1975; Dejonghe 1986).
But this simplicity is achieved at the cost of having to assume the forms 
of $\Phi(\varpi,z)$ and $f_-$ at the outset.
It was shown above that the potential follows uniquely from 
the observed $\overline{v}_{p}(X,Z)$ and $\overline{v_p^2}(X,Z)$ 
in a two-integral, edge-on system -- one is not free to assume 
arbitrary functional forms for $\Phi$.
Furthermore the odd part of $f$ can be derived in a 
model-independent way from the observed rotational velocity 
field.
It follows that the approach taken by these authors is overdetermined.
Models constructed in an assumed potential are unlikely to reproduce a 
two-dimensional set of velocity dispersions, since the information 
contained in those data was not used.
And if the model should fail to reproduce the kinematical data, the failure 
could be due to the breakdown of either of two assumptions: that 
mass follows light, or that $f$ depends on only two integrals.
In the method described here, on the other hand, the potential is
constructed to have the unique form that is consistent with the two-integral
assumption for $f$ and with its lowest observable moments.
A failure of the model to reproduce any additional kinematical data
could only mean that $f\ne f(E,L_z)$.

The model-building technique has nevertheless been successfully applied 
to a few galaxies, notably M32 (van der Marel et al. 1994;
Dehnen 1995; Qian et al. 1995).
But this apparent success should be interpreted with caution.
It is possible that these galaxies satisfy all of the assumptions 
made in the model building, i.e. axisymmetry, $f=f(E,L_z)$, and $M/L =$ 
constant.
But it is also possible that the published models would fail to reproduce 
a full, two-dimensional set of velocities if such data were 
available.
Future observations of these galaxies should reveal which of these 
two possibilities is correct.

Of course, Binney, Davies \& Illingworth (1990) were concerned with 
situations where the kinematical data are confined to just a few cuts 
across the image of the galaxy.
Given such limited data, their approach seems entirely justified.
At the same time, our discussion makes clear that -- when two-dimensional
data {\it are} available, as is increasingly often the case -- 
one can and should infer the kinematics and potential directly from the 
velocities.

Merrifield (1991) proposed a test for the sufficiency of two-integral 
models when describing edge-on, axisymmetric galaxies.
His test is based on the assumption that mass follows light, and uses as
a discriminant the measured velocity dispersions.
Since one can always construct the potential so as to reproduce 
the velocity dispersion data in a two-integral system, as shown here, 
conclusions based on Merrifield's test are contingent on the 
assumption that mass follows light.

\section{PRACTICAL INVERSION ALGORITHMS}

The formal solutions given above to the axisymmetric inverse problem are of
little use when dealing with real data, since the inversion equations are
ill-conditioned in the sense understood by statisticians (Miller 1974;
O'Sullivan 1986).
This means that errors or incompleteness in the data will be amplified
when going from data space to model space unless some objective
smoothness condition is placed on the solution.
The degree to which errors are amplified depends on the number of effective 
differentiations that separate the model from the data. 
For instance, the distribution function $f(E,L_z)$ is essentially a second 
derivative of the number density (Eq. \ref{fela}) which is itself a 
half-order derivative of the surface brightness (Eq. \ref{nur}) -- hence
the determination of $f$ is strongly ill-conditioned even though 
formally unique.
Some authors have ignored this element of the problem and carried 
out direct deprojections without imposing smoothness constraints.
The results tend to be spectacularly noisy (e.g. Figure 2 of Kuijken 1995).
Another approach, equally ill-advised, is to replace the data by ad
hoc fitting functions for which the inversions can be carried out 
exactly (e.g. Qian et al. 1995).
Unless these smooth functions are generated from the data using a 
nonparametric algorithm, they will contain a bias that is likewise 
amplified by the inversion.
Solutions obtained in this way may look appealing in a $\chi^2$ sense 
but are likely to be far from the true solutions.

Richardson-Lucy iteration (e.g. Dehnen 1995) is genuinely nonparametric,
going in discrete steps from an initial guess to the (ill-defined) solution 
that would have been obtained via direct inversion of the data.
However the success of this technique depends on the accuracy of 
the initial guess. 
A bad guess will require many iterations before the projected solution 
approximates the data, at which point the solution is likely to be 
unacceptably noisy.
One typically halts the iteration before this occurs but the solution 
so obtained will accordingly be biased in the direction of the 
initial guess by some unknown amount.
Furthermore, solutions obtained in this way tend to achieve their 
optimal degree of smoothness in one part of solution space while 
remaining undersmoothed in other parts.

Our goal is an algorithm which generates solutions that are 
characterized by acceptably small values of {\it both} the noise and the 
bias.
Modern nonparametric methods (M\"uller 1980; H\"ardle 1990; Wahba 1990; 
Scott 1992; Green \& Silverman 1994) are designed with precisely 
this goal in mind.
These methods make no assumptions about the global form of the 
solutions; they deal with the ill-conditioning by imposing smoothness 
via an effectively local constraint on the level of fluctuations in the 
solution, an approach called ``regularization.''
They thus avoid the pitfalls of both Richardson-Lucy iteration (no 
regularization) and model fitting (parametric bias).

Our focus is on situations where the velocities are measured discretely,
as would be the case for stars in a globular cluster, planetary nebulae
around a galaxy, galaxies in a cluster, etc.
Modifying the algorithms to deal with spatially-continuous data 
is straightforward.

We assume that the number density profile $\nu(\varpi,z)$ of the
kinematical sample is known.
Nonparametric algorithms for estimating $\nu$ in the spherical geometry 
have been described by Merritt (1993a) in the case that the surface 
brightness is measured directly, and by Merritt \& Tremblay (1994) 
in the case that the surface brightness must first be inferred 
from discrete positions.
These algorithms are easily generalized to the 
axisymmetric case.
We also continue to assume, as discussed above, that the observer lies in the 
equatorial plane of the observed system, since otherwise the 
inverse problem is guaranteed not to have a unique solution 
(Gerhard \& Binney 1996).

\subsection{A Variational Problem}

The inverse problems to be solved here belong to a class of problems that
have been well studied by statisticians in the last few years 
(e.g. Wahba 1990).
Let the data be $d_1, d_2, ..., d_n$ and the model $u$.
The data are related to the model via
\begin{equation}
d_i = {\cal L}_iu + e_i,
\end{equation}
where the ${\cal L}_i$ are known linear functionals -- in our case,
projections -- of $u$, and the $e_i$ represent measurement errors or scatter
from some other source.
For instance, in the determination of the rotational velocity field
$\overline{v}_{\phi}$, we would have
\begin{equation}
u = \overline{v}_{\phi}(\varpi,z)
\end{equation}
and
\begin{eqnarray}
{\cal L}_iu & = & 2X_i\Sigma(X_i,Z_i)^{-1}\int_{X_i}^{\infty}\nu(\varpi,Z_i) u
(\varpi,Z_i) {d\varpi\over\sqrt{ \varpi^2 - X_i^2}} \nonumber \\
& = & \overline{v}_p(X_i,Z_i), 
\end{eqnarray}
the observed rotation at point $(X_i,Z_i)$.
If the line-of-sight rotational velocity were measured directly, then
$d_i=\overline{v}_p(X_i,Z_i)$ and the $e_i$ would be measurement errors.
If instead discrete velocities were measured, then $d_i = v_i$, the observed
velocity of the $i$th star, and $e_i=v_i-\overline{v}_p(X_i,Z_i)$, the
deviation (due both to measurement errors and to the intrinsic 
velocity dispersion) of the $i$th star's measured velocity from the 
true mean line-of-sight velocity at point $(X_i,Z_i)$.

We seek a smooth function $\hat u$ such that ${\cal L}\hat u$ is suitably
close to the data.
One possible approach is to construct a continuous approximation to the
function represented by the data -- e.g. $\overline{v}_p(X,Z)$ --  using a
kernel or spline smoother, and then to operate mathematically on this
function via the inverse operator ${\cal L}^{-1}$ (if it exists) to produce
the estimate $\hat{u}$.
This approach is consistent in the sense defined by statisticians
as long as the functions that are fit to the data are asymptotically 
unbiased, i.e. nonparametric (Wahba 1990, p. 19), and such an 
approach has been used with success by Merritt \& Gebhardt (1994) to 
solve the dynamical inverse problem in spherical geometry.
Such an approach will not be followed here because one of the inverse problems
that we wish to solve -- the recovery of $f(E,L_z)$ from $\nu$,
$\overline{v}_{\phi}$ and $\Phi$ -- does not have a simple inverse operator.
In addition, one sometimes wishes to place physical constraints (e.g. 
positivity) on the solutions and this is extremely difficult to do if the
function approximations are carried out in data space rather than solution
space.

Instead we define our solutions implicitly, as the functions $\hat u$ that
minimize functionals of the form
\begin{equation}
n^{-1}\sum_{i=1}^n\left({\cal L}_iu - d_i\right)^2 + \lambda J(u). 
\label{opt}
\end{equation}
The first term in equation (\ref{opt}) measures the fidelity of the 
model to the data.
Minimizing this term alone would lead to an unacceptably noisy 
solution due to the ill-conditioning of the inverse operator.
The second term measures the degree to which the solution is unsmooth.
One standard choice for the ``penalty function'' $J(u)$ is
\begin{equation}
J(u)=J_m(u) =
\sum_{\alpha+\beta=m}{m!\over\alpha!\beta!}\int\int\left({\partial^mu\over
\partial x^{\alpha}\partial y^{\beta}}\right)^2 dx dy, \ \ \ \ m\ge 2.
\label{penaltya}
\end{equation}
(Here and below, a two-dimensional $u=u(x,y)$ is assumed.)
The function $\hat u$ that minimizes equation (\ref{opt}) with penalty function
(\ref{penaltya}) is the so-called ``thin plate spline'' solution (Duchon 1977;
Meinguet 1979).
Remarkably, this solution can be found in essentially analytic form for
many simple operators ${\cal L}$ (Wahba \& Wendelberger 1980).

The qualitative nature of the solution to equations (\ref{opt}) and 
(\ref{penaltya})
depends on the choices of $\lambda$ and $m$.
The penalty function acts as a
low-pass filter, with $\lambda$ controlling the half-power point of the
filter and $m$ the steepness of the roll-off (Craven \& Wahba 1979).
Typically one chooses $m=2$; the smoothness of the solution is then
controlled by varying $\lambda$.
Too large a value for $\lambda$ will oversmooth the solution, i.e. reduce its
curvature; too small a value will yield an overly noisy solution.
The larger the data set and the smaller the measurement errors, the smaller
the value of $\lambda$ that can be profitably used and the more accurate the
solution.
Thus the bias -- the average deviation of the solution from the true 
function -- falls to zero as the sample size $n$ is increased.
In parametric algorithms, by contrast, the bias remains finite even as 
$n$ tends to infinity, since the adopted parametric form is
guaranteed to be incorrect at some level and this incorrectness does not 
diminish as $n$ is increased.
It is for this reason that parametric methods are unsuited to 
data-based inverse problems.

Data are typically incomplete or cover only a limited region.
The use of a penalty function guarantees that the solution will tend toward a
predictable form in parts of solution space that are not strongly constrained
by the data.
For instance, the thin-plate spline solution with $m=2$ yields a solution
that is linear in $x$ and $y$ wherever the data do not force a different
functional form, since any linear function is assigned zero penalty by
expression (\ref{penaltya}).
This very useful feature of solutions found via penalty functions means that
there is never any need to extend or augment the data in ad hoc 
ways.
Nonparametric techniques based on direct smoothing of the data, 
e.g. via kernels, do not share this nice property.

Many variations on this basic scheme exist.
If the data errors are not normally distributed, one can choose the first
term in equation (\ref{opt}) to be some more robust measure of the fidelity of the
model to the data.
Such a modification would be appropriate in the
study of galaxy cluster dynamics where interlopers are common.
Another variation is to modify the form of the penalty function.
The simplest such variation is to add a factor $k^{m-\alpha}$ with $k$ some
positive constant; changing $k$ is equivalent to changing the units in which
$y$ is measured.
One can also include in the penalty function a weighting function $w(x,y)$
that is large where the solution $u$ is expected to be small, thus
encouraging the fluctuations in the solution to remain a fixed fraction of
$u$ in amplitude everywhere.
Finally, one can replace the differential operator $\partial^m\over\partial
x^{\alpha}\partial y^{\beta}$ by some more
complicated operator, which defines as ``perfectly smooth'' some different
class of functions (e.g. Silverman 1982).

For many choices of $J(u)$, including the thin plate penalty function
defined above, the quantity to be minimized is quadratic in $u$.
This means that a standard quadratic programming routine (Rustagi 1994) 
can be used to find the solution on some grid in model space.
The use of a quadratic programming algorithm also allows the easy imposition
of linear constraints, such as positivity, on the solution if desired.

There is a huge literature concerning the choice of the smoothing 
parameter $\lambda$.
While ``objective'' schemes exist, a standard practice is to 
slowly decrease $\lambda$ until the solution begins to exhibit 
small-scale fluctuations, then stop.

\subsection{The Rotational Velocity Field}

Consider first the recovery of the rotational velocity field in the 
meridional plane $\overline{v}_{\phi}(\varpi,z)$.
We seek the function $\hat{\overline{v}}_{\phi}$ that minimizes
\begin{equation}
\chi^2=\sum_{i=1}^n \left({v}_i - {\cal L}_i {\overline v_{\phi}}\right)^2 + 
\lambda\int\int\left[\left({\partial^2{\overline
v}_{\phi}\over \partial\varpi^2}\right)^2 + 2\left({\partial^2{\overline
v}_{\phi}\over \partial\varpi\partial z}\right)^2 +
\left({\partial^2{\overline v}_{\phi}\over \partial z^2}\right)^2\right]
d\varpi dz, 
\label{opta}
\end{equation}
with $v_i$ the measured velocities and
\begin{equation}
{\cal L}_i{\overline v_{\phi}} =
2X_i\Sigma(X_i,Z_i)^{-1}\int_{X_i}^{\infty}\nu(\varpi,Z_i){\overline{v}}_{\phi
}(\varpi,Z_i) {d\varpi\over\sqrt{\varpi^2-X_i^2}}.
\label{proja}
\end{equation}
We choose $\overline{v}_{\phi}$ as our function to be optimized
rather than the physically more natural choice $\nu\overline{v}_{\phi}$, 
since $\overline{v}_{\phi}$ is expected to be a more nearly 
linear function of the coordinates and hence better suited to a 
second-derivative penalty function. 

We begin by defining a Cartesian grid in model space, $\varpi_j = \varpi_1 
+ (j-1)\delta\varpi$, $1\le j\le N$, and $z_k = z_1 + (k-1)\delta 
z$, $1\le k\le N$.
The variables to be solved for are the $N^2$ values 
$v_{j,k}$ of the rotational velocity on the grid.
The derivatives and integrals on the right hand side of (\ref{opta}) can 
easily be represented as linear combinations of the 
$v_{j,k}$ using standard expressions from 
numerical analysis; for details from a very similar problem, see 
Merritt \& Tremblay (1994).
Placing the $v_{j,k}$ into a linear 
vector ${\bf g}$, one can then write
\begin{equation}
\chi^2 = {\bf g}^T{\bf A} {\bf g} - {\bf B}{\bf g} + C
\end {equation}
where the quantities ${\bf A}$, ${\bf B}$ and $C$ are 
straightforward to compute but messy and will not be given here.
The resulting expression is quadratic in the unknowns, 
and the constraints $v_{j,k} \ge 0$ are linear.
The values ${\hat v}_{j,k}$ that minimize 
$\chi^2$ can accordingly be found via quadratic programming.

Figure 1 shows an estimate of ${\overline v_{\phi}}(\varpi,z)$ 
derived using
5000 discrete velocities drawn randomly from an oblate model belonging 
to Lynden-Bell's (1962) family.
The adopted model is the flattest one consistent with non-negative $f$,
having $b/a=0.638$ in Lynden-Bell's notation and $a=-0.814$ in 
the notation of Dejonghe (1986).
(These models have non-constant ellipticity, becoming spherical at large
radii for all values of $b/a$.)
The odd part of $f$, which specifies the degree of ordered motion, was
set to the function derived in the appendix of Lynden-Bell's paper when 
generating the Monte-Carlo data. 
The resultant rotational velocity field has 
$\sigma_{\varpi}=\sigma_{\phi}=\sigma_z$ everywhere, an 
``oblate isotropic rotator.''
Note that even this extreme model from Lynden-Bell's sequence is only mildly
flattened, except very near the center, and the rotation is correspondingly
small; the peak rotation velocity is only about one half of the central,
one-dimensional velocity dispersion.
In spite of the small signal, however, the inversion algorithm recovers 
the rotational velocity field reasonably well near the center; the main 
defect is a too-shallow rise of the rotation velocity near $\varpi=0$, an 
inevitable result of the smoothing.
This example suggests that data sets consisting of a few thousand velocities 
are none too large for recovering the rotational velocity field 
of an axisymmetric system in a model-independent way.

\subsection{The Velocity Dispersions and Potential}

We now wish to make estimates of $\sigma(\varpi,z)$ and 
$\sigma_{\phi}(\varpi,z)$.
We search for the functions $\hat{\sigma}^2$ and 
$\hat{\overline{v^2_{\phi}}}$ that minimize
\begin{eqnarray}
\chi^2 & = & \sum_{i=1}^n \left[{v}_i^2 - {\cal L}_i 
\left\{ \sigma^2, \overline{v^2_{\phi}}\right\} \right]^2 \nonumber \\
& + & \lambda\int\int\left[\left({\partial^2 \sigma^2\over \partial\varpi^2}
\right)^2 + 2\left({\partial^2 \sigma^2 \over \partial\varpi\partial z}
\right)^2 + \left({\partial^2 \sigma^2 \over \partial z^2}\right)^2\right]
d\varpi dz \nonumber \\
& + & \lambda\int\int\left[\left({\partial^2\overline{v^2_{\phi}}\over \partial
\varpi^2}\right)^2 + 2\left({\partial^2 \overline{v^2_{\phi}}\over\partial\varpi
\partial z}\right)^2 + \left({\partial^2 \overline{v^2_{\phi}} \over \partial z^2}
\right)^2\right] d\varpi dz
\label{optb}
\end{eqnarray}
subject to the constraints of equation (\ref{magic}):
\begin{equation}
{\partial\nu\over\partial\varpi}{\partial\sigma^2\over\partial z} - 
{\partial\nu\over\partial z}{\partial\sigma^2\over\partial\varpi} +
{\nu\over\varpi}{\partial\over\partial z}\left(\sigma^2 - 
\overline{v^2_{\phi}}\right) = 0
\label{magic3}
\end{equation}
as well as the usual positivity constraints.
The operator ${\cal L}_i$ is now defined as
\begin{equation}
{\cal L}_i\left\{\sigma^2,\overline{v^2_{\phi}}\right\} = 
2\Sigma(X_i,Z_i)^{-1}\int_{X_i}^{\infty}\nu(\varpi,Z_i)\left[\left( 
1-{X_i^2\over \varpi^2}\right)\sigma^2(\varpi,Z_i) + 
{X_i^2\over\varpi^2}\overline{v^2_{\phi}}(\varpi,Z_i)\right]
{d\varpi\over\sqrt{\varpi^2-X_i^2}}.
\label{projb}
\end{equation}
If the velocities contain measurement errors, these 
should be subtracted in quadrature from the $v_i$.

We define a double grid in $\{\varpi, z\}$, with the first $N^2$ 
points containing the $\sigma^2$ values and the second 
$N^2$ points the $\overline{v^2_{\phi}}$ values.
The constraints (\ref{magic3}) are approximated via finite-difference 
derivatives, which have the effect of linking together the values on the 
two grids.
Finally, one obtains $\sigma_{\phi}$ via $\sigma_{\phi}^2 = 
\overline{v^2_{\phi}} - \overline{v}^2_{\phi}$, with $\overline{v}^2_{\phi}$
having been recovered in the previous step.

Before applying the algorithm to pseudo-data, one would like some 
assurance that the information contained in the observed 
mean-square velocity field $\overline{v_p^2}(X,Z)$
is sufficient, in principle, to uniquely constrain 
both $\sigma(\varpi,z)$ and $\sigma_{\phi}(\varpi,z)$.
Ideally this hypothesis would be demonstrated by finding
analytical expressions for the inverse relations.
Such expressions could not be found (which may simply reflect 
limitations in the author's mathematical abilities), and instead 
the case will here be made numerically.

Figure 2 shows how the integrated square error in the solutions 
$\hat{\sigma}^2(\varpi,z)$, $\hat{\sigma}^2_{\phi}(\varpi,z)$,
\begin{equation}
{\rm ISE} = 
\int\int\left(\sigma^2-\hat{\sigma}^2\right)^2d\varpi dz + 
\int\int\left(\sigma^2_{\phi}-\hat{\sigma}^2_{\phi}\right)^2d\varpi dz,
\label{ISE}
\end{equation}
depends on the degree to which the constraints (\ref{magic3}) are 
imposed.
That is, these constraints were replaced by
\begin{equation}
-\Delta \le {\partial\nu\over\partial\varpi}{\partial\sigma^2\over\partial z} - 
{\partial\nu\over\partial z}{\partial\sigma^2\over\partial\varpi} +
{\nu\over\varpi}{\partial\over\partial z}\left(\sigma^2 - 
\overline{v^2_{\phi}}\right) \le \Delta,
\label{magic4}
\end{equation}
with the parameter $\Delta>0$ -- a constant at all points on the 
solution grid -- determining how stringently the contraints are imposed.
The algorithm was given as input the exact values, on a grid, of the mean 
square velocity $\overline{v_p^2}(X,Z)$ of the rotating oblate model 
of Figure 1.
Solutions were then obtained as a function of $\Delta$.
Figure 2 demonstrates that the error in the solutions tends to zero
along with $\Delta$, which suggests -- at least for this particular 
model -- that the constraints (\ref{magic3}) contain just enough 
information to yield the correct inversion.
This experiment was repeated for a number of other models from the 
Lynden-Bell sequence of models.
For each model examined, the solution with $\Delta=0$ was found to be
extremely close to the true solution.
These experiments hardly constitute a rigorous proof that the inversion is 
uniquely defined for every possible two-integral oblate model, however.

Figures 3a and 3b illustrate the recovery of $\sigma(\varpi,z)$ and 
$\sigma_{\phi}(\varpi,z)$ given the same 5000 velocities drawn from the
rotating oblate model of Figure 1.
Given the uniqueness, just demonstrated, of the inversion for this model,
the discrepancies apparent in the numerical solutions of Figure 3
are primarily due to the finiteness of the data set rather than 
to any limitations of the algorithm.
Different, 5000-velocity Monte-Carlo realizations of the same model 
were found to yield solutions that were different in their details but
that had roughly the same average error.

The gravitational potential is obtained by applying either 
of equations (10) or (11), in finite-difference form, to the 
derived $\sigma$ and $\sigma_{\phi}$.
The result is shown in Figures 3c and 3d.
The derived potential is remarkably similar to the correct one 
except for slightly reduced gradients, again an unavoidable
result of the smoothing.

\subsection {The Distribution Function}

We now want to find the functions $f_+(E,L_z)$ and $f_-(E,L_z)$ that generate 
the estimated number density $\hat{\nu}(\varpi,z)$ and rotational 
velocity field $\hat{\overline v}_{\phi}(\varpi,z)$ in the reconstructed 
potential $\hat{\Phi}(\varpi,z)$.
This is a simpler version of a problem solved in Merritt (1993b), 
and we follow a similar approach here.
A rectangular grid is defined in $(E,L_z)$-space, and the 
unknowns ${f_+}_{j,k}$ and ${f_-}_{j,k}$ are assumed constant over each cell 
(Figure 4).
The configuration-space density at any point $(\varpi,z)$ is the 
sum of the contributions $\nu_{cell}=4\pi\varpi^{-1}\Delta E\Delta L_z 
{f_+}_{j,k}$, for each cell such that ${L_z}_{j,k} < 
\varpi\sqrt{2\left[E-\Phi(\varpi,z)\right]}$.
(Cells that lie partly below and partly above this curve 
contribute by an amount proportional to the area that lies 
below the curve.)
Similarly, each cell contributes $\nu{\overline{v}_{\phi}}_{cell}
= 4\pi\varpi^{-2}
\Delta E \Delta L_z^2
{f_-}_{j,k}$ to the estimate of 
$\nu(\varpi,z)\overline{v}_{\phi}$.
The quantity to be minimized is
\begin{eqnarray}
\chi^2 & = & \sum_{i=1}^n \left({\nu}_i - \sum_{cell}\nu_{cell}\right)^2 + 
\nonumber \\ 
& & \int\int\left[\lambda_1(E)\left({\partial^2 f_+ \over \partial E^2}\right)^2 
 + 2\sqrt{\lambda_1(E)\lambda_2(E)} \left({\partial^2 f_+\over  
\partial E\partial L_z}\right)^2 
+ \lambda_2(E)\left({\partial^2 f_+\over \partial L_z^2} 
\right)^2\right ] dE dL_z, 
\label{optc}
\end{eqnarray}
in the case of $f_+$, 
and
\begin{eqnarray}
\chi^2 & = & \sum_{i=1}^n \left({\nu}_i{\overline{v}_{\phi}}_i - 
\sum_{cell}\nu{\overline{v}_{\phi}}_{cell}\right)^2 + \nonumber \\ 
& & \int\int \left[\lambda_1(E)\left({\partial^2 f_- \over \partial 
E^2}\right)^2 
 + 2\sqrt{\lambda_1(E)\lambda_2(E)}\left({\partial^2 f_-\over  
\partial E\partial L_z}\right)^2 
+ \lambda_2(E)\left({\partial^2 f_-\over \partial L_z^2} 
\right)^2\right] dE dL_z, 
\label{optd}
\end{eqnarray}
in the case of $f_-$.
The integrals are understood to extend over the region in the 
$(E,L_z)$ plane such that $0<L_z<L_{max}(E)$, where $L_{max}(E)$ is 
the maximum value of $L_z$ attainable at energy $E$ (Fig. 4).
The smoothing parameters $\lambda_1$ and $\lambda_2$ 
are allowed to depend on the energy since $f$ falls off much more rapidly 
than linearly with $E$, and a penalty function with constant 
$\lambda$ would not strongly penalize a solution that was smooth 
at large $|E|$ but noisy for $E\approx 0$.
Furthermore, a different degree of smoothing is permitted in the 
$E$ and $L_z$ directions since $f$ is expected to be much more 
homogeneous in $L_z$ than in $E$.

Figure 5 illustrates the recovery of $f_+$ and $f_-$ using the 
approximations to $\Phi(\varpi,z)$ and $\overline{v}_{\phi}$ obtained 
in the previous sections.
The smoothing parameters were chosen to vary as $|E|^{-1.5}$, 
with $\lambda_2/\lambda_1=100$.
Once again, the only obvious systematic error takes the form of a bias 
due to the smoothing.

The derivation of $f(E,L_z)$ given $\nu(\varpi,z)$ and $\Phi(\varpi, 
z)$ is a classical problem in stellar dynamics.
Lynden-Bell (1962), Hunter (1975) and Dejonghe (1986) 
demonstrated the uniqueness of the solution, and recently Hunter \& 
Qian (1993) have presented an elegant algorithm for carrying 
out the inversion given analytic expressions for $\nu$ and 
$\Phi$.

Hunter \& Qian's algorithm is not regularized and hence not 
suitable for use with real data.
It has nevertheless been applied in this way, after first replacing the 
data with analytic models (Qian et al. 1995).
This approach is dangerous at best since any slight error in the 
choice of fitting functions will be amplified enormously by the 
inversion operator.
Solutions so obtained for $f$ may be smooth and physically appealing, 
but they are unlikely to be close to the true solutions.

From a numerical point of view, the only non-trivial aspect of 
solving the Lynden-Bell--Hunter problem 
is the regularization; the inversion itself is essentially just a 
matrix operation once the integral operator has been represented via 
some discrete approximation.
Much effort has been directed 
recently toward refining the discretization (Kuijken 1995) 
or the inversion (Hunter \& Qian 1993), with curiously little 
attention paid to the more crucial question of regularization.

\section{SUMMARY}

Given measurements of the line-of-sight rotational velocity and 
the velocity dispersion over the two-dimensional image of an edge-on, 
axisymmetric galaxy, one can derive the unique functions 
$\Phi(\varpi,z)$ and $f(E,L_z)$ that reproduce those data, 
without the necessity of assuming that mass follows light.
The validity of such a model can then be tested using other data, 
such as the full line-of-sight velocity distributions or the proper 
motions.
Failure to reproduce the data using this method would imply that $f$ 
depends on a third integral of motion (or that one of the geometric 
assumptions, axisymmetry or zero inclination, was violated).
This approach is a generalization of the one pioneered by Binney, 
Davies and Illingworth (1990) which assumes that mass follows 
light and which derives both $\Phi$ and $f$ from the luminosity 
distribution alone, without using the kinematical data
except in the normalization of $\Phi$.
We have presented numerical algorithms, suitable for use with noisy and 
incomplete data, for carrying out the required inversions and shown that 
they provide smooth and accurate estimates of $f$ and $\Phi$ using 
simulated data sets derived from oblate models.

In a companion paper, the algorithms described here will be applied to
velocity data from the globular cluster $\omega$ Centauri.

The results described here suggest the following avenues for future work.

1. A mathematical -- as opposed to numerical -- demonstration of uniqueness
(or non-uniqueness) of the solutions to the coupled equations (\ref{magic})
and (\ref{disp}) would be of fundamental importance.

2. The second-derivative penalty functions adopted here are almost certainly
non-optimal and it would be worthwhile to investigate more general forms.
In addition, it is possible that the solutions to the roughness-penalized
inverse problems defined here can be found analytically, 
as in Wahba \& Wendelberger (1980).

3. As shown most recently by Gerhard \& Binney (1996),
the axisymmetric inverse problem becomes 
under-determined when the stellar system is viewed from a 
direction that is not parallel to the symmetry plane.
Additional information must therefore be added if one is to solve
the inverse problem for an arbitrary orientation.
(One would also like to {\it infer} that orientation from the data.)
There is at present no clear understanding of what form that extra 
information should take.

4. One can define an alternate inverse problem based on 
the {\it Ansatz} that mass follows light, or more generally 
that the form of the potential $\Phi(\varpi,z)$ is known a priori; 
for instance, the potential might be derived from X-ray data, 
from the statistics of gravitational lensing, etc.
This is of course the usual assumption that is made when 
modelling axisymmetric galaxies; however the goal would be to 
learn something unique about the degree of velocity anisotropy 
$\sigma_{\varpi}/\sigma_z$ from the velocity moment data,
as in the analogous spherical problem (Binney \& Mamon 1982).
Unfortunately one can show that the assumption of a known potential 
is {\it not} sufficient in itself to uniquely constrain the 
anisotropy, even for an edge-on system; one must make additional 
assumptions, e.g. that the velocity ellipsoid is everywhere 
aligned with the coordinate axes.
Nevertheless, more work along these lines would help to elucidate 
the more difficult inverse problem for a three-integral $f$.

\bigskip

This work was supported by NSF grant AST 90-16515 and NASA grant 
NAG 5-2803.
T. Fridman kindly assisted in the plotting of the figures.
I thank W. Dehnen and O. Gerhard for lively discussions on the topic
of axisymmetric modelling, and for pointing out some mathematical 
errors in an early version of the paper.
I also thank H. Dejonghe for many useful insights on problems of this sort.

\clearpage

\figcaption[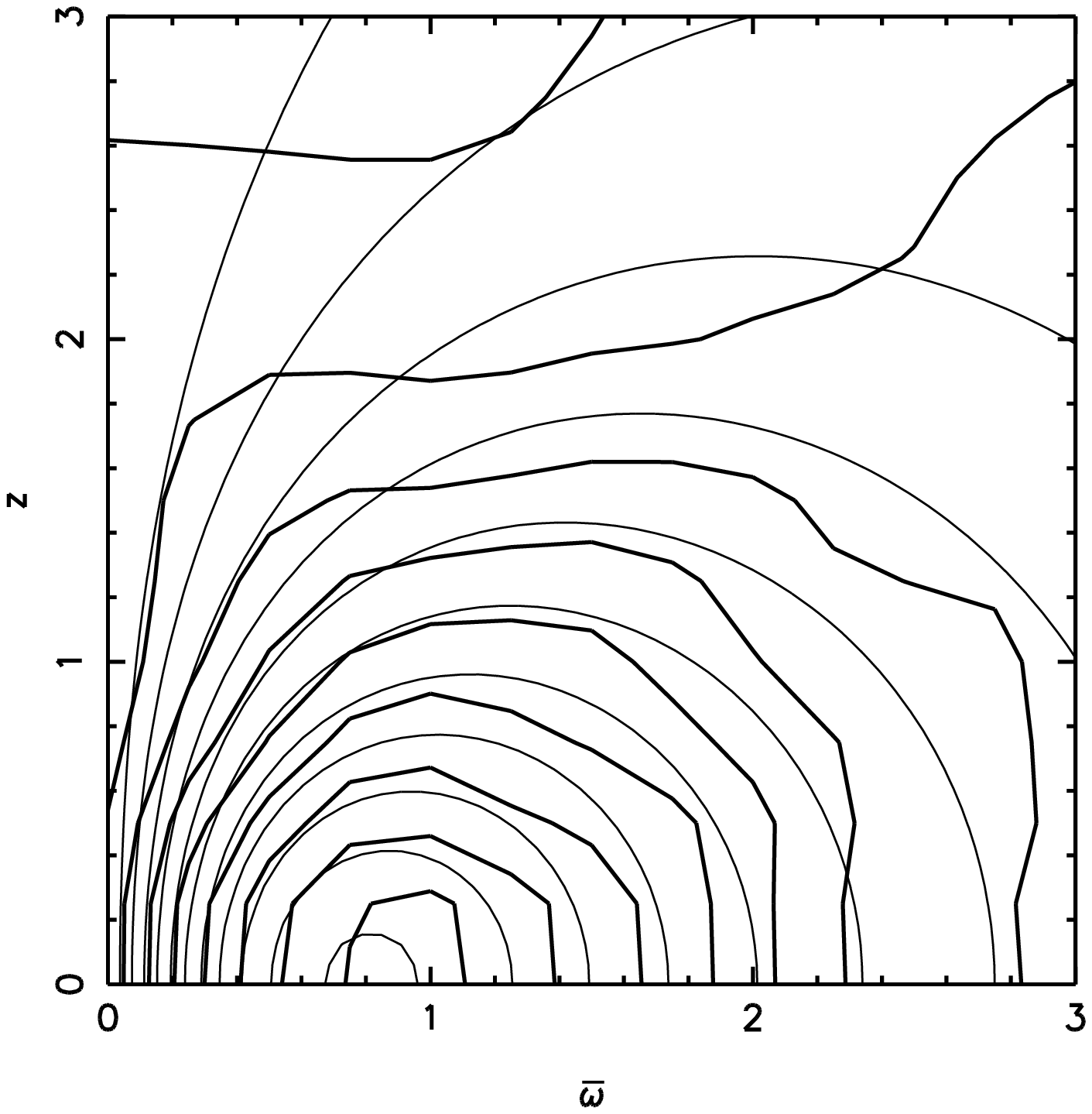]{\label{fig1}} 
Recovery of the rotational velocity field $\overline{v}_{\phi}(\varpi,z)$ 
given 5000 velocities derived from an oblate, Lynden-Bell (1962) 
model with maximum flattening.
Heavy contours: estimated $\overline{v}_{\phi}$; thin contours: 
exact $\overline{v}_{\phi}$.

\figcaption[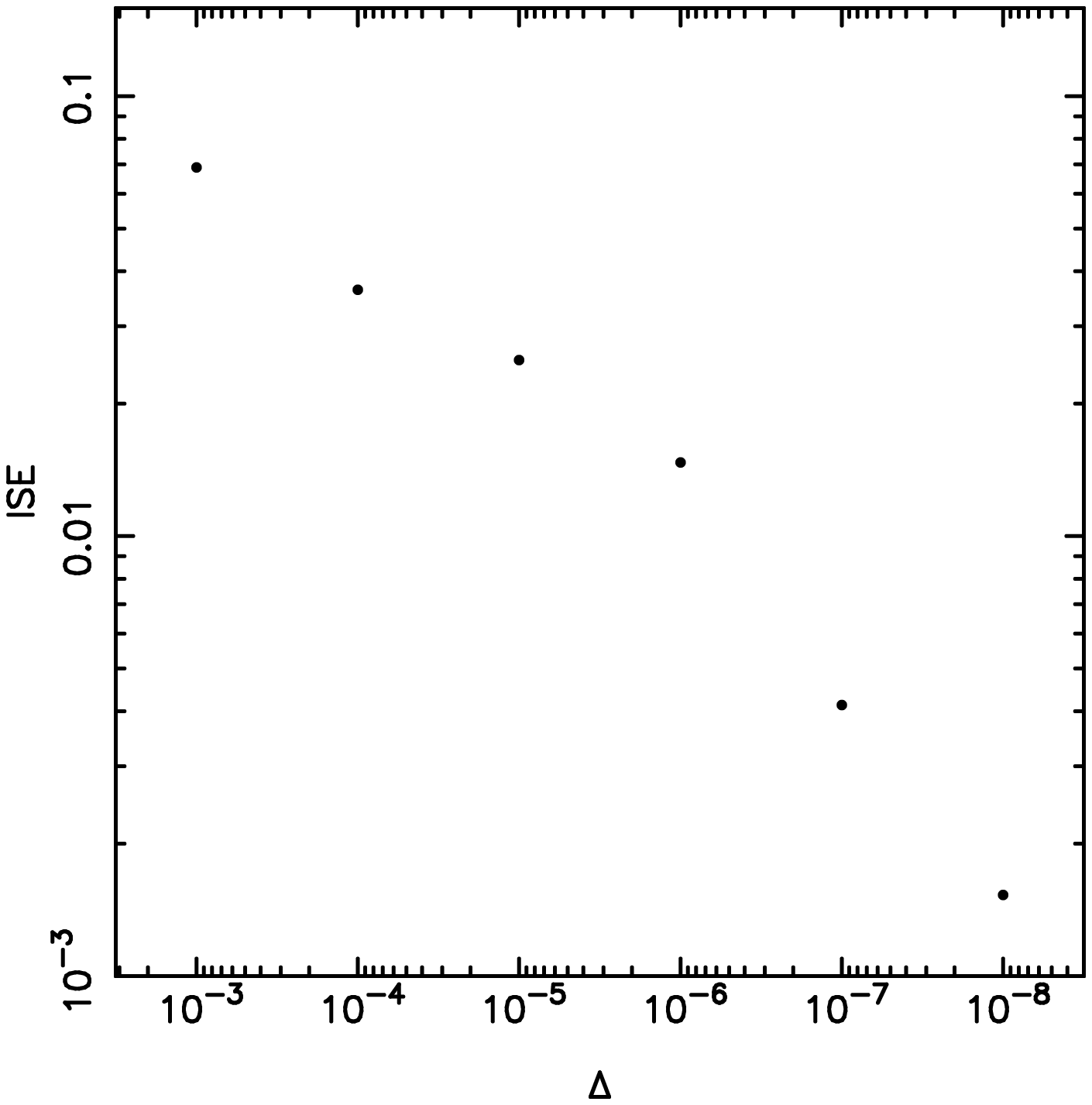]{\label{fig2}}
Dependence of the ISE of the velocity dispersion solutions, Eq. 
(\ref{ISE}), on the parameter $\Delta$ that determines how 
stringently the constraints (\ref{magic4}) are applied.

\figcaption[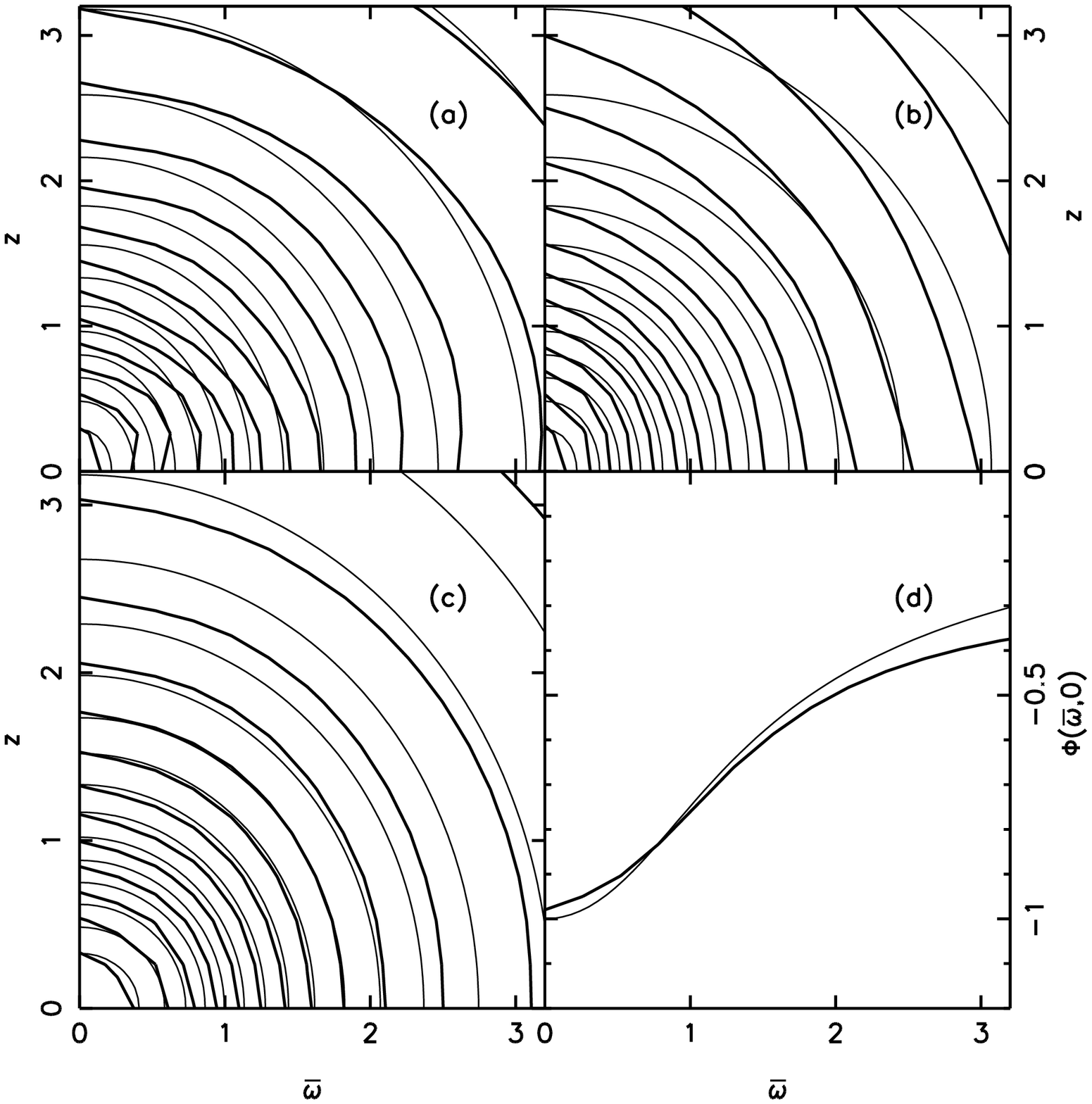]{\label{fig3}} 
Recovery of the velocity dispersions and potential from the same 
data used Fig. 1.
Heavy contours: estimated values; thin contours: exact values.
(a) $\sigma$; (b) $\sigma_{\phi}$; (c) 
$\Phi$; (d) $\Phi(\varpi,0)$.

\figcaption[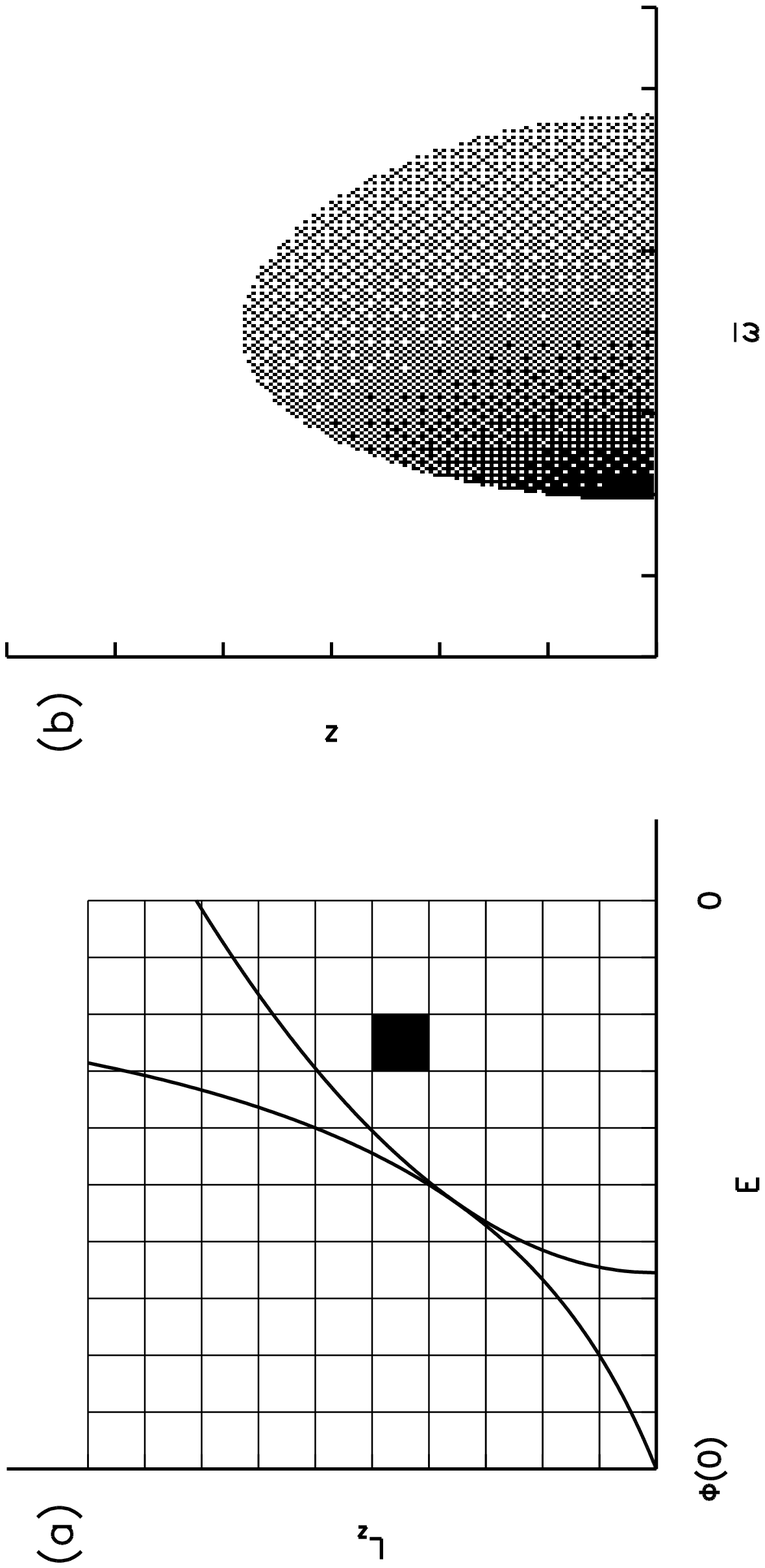]{\label{fig4}} 
Grid-based scheme for recovery of $f_+(E,L_z)$.
(a) Solution grid.
Upper curve is the maximum $L_z$ as a function of $E$; lower curve is 
the characteristic curve for a given $(\varpi,z)$.
(b) Projection of a single $(E,L_z)$ grid cell into $(\varpi,z)$-space.

\figcaption[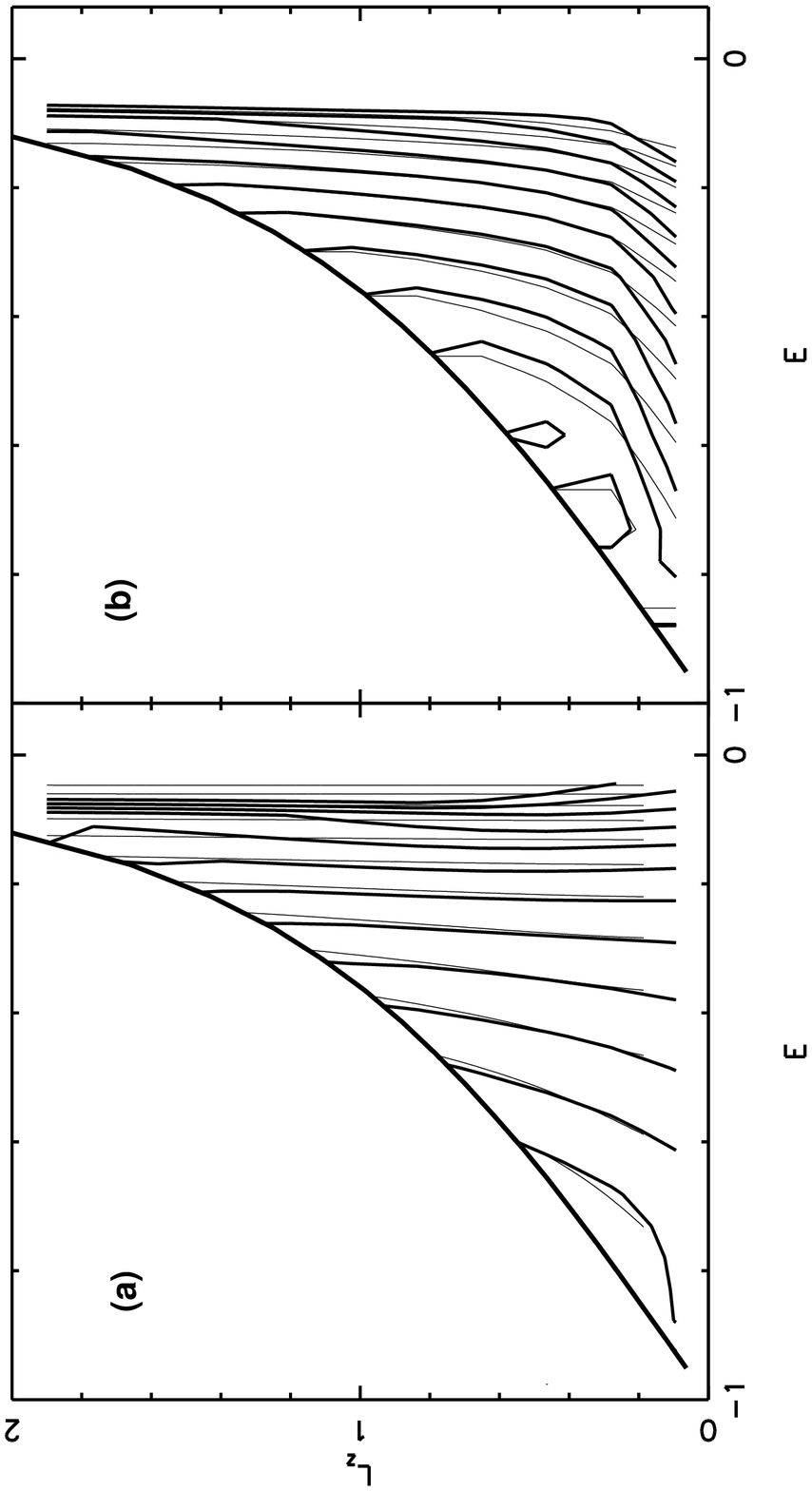]{\label{fig5}} 
Recovery of $f_+(E,L_z)$ (a) and $f_-(E,L_z)$ (b) given data generated 
from the oblate Lynden-Bell model.
Heavy contours: estimated values; thin contours: exact values. 

\setcounter{figure}{0}

\begin{figure}
\plotone{figure1.ps}
\caption{ }
\end{figure}

\begin{figure}
\plotone{figure2.ps}
\caption{ }
\end{figure}

\begin{figure}
\plotone{figure3.ps}
\caption{ }
\end{figure}

\begin{figure}
\plotone{figure4.ps}
\caption{ }
\end{figure}
  
\begin{figure}
\plotone{figure5.ps}
\caption{ }
\end{figure}
  
\end {document}